\documentclass[12pt]{article}

\usepackage{moreverb}

\usepackage[colorlinks,bookmarksopen,bookmarksnumbered,citecolor=blue,urlcolor=red]{hyperref}

\usepackage{natbib}

\usepackage{amsmath,amssymb,algorithm,algpseudocode,bm,tikz,enumitem,mathabx,multicol,fullpage}
\usetikzlibrary{shapes,backgrounds}

\begin{document}

\title{Accelerated Nonparametrics for Cascades of Poisson Processes}

\author{Chris. J. Oates\footnote{E-mail: christopher.oates@uts.edu.au. This work was supported by UK EPSRC EP/E501311/1 and by the ARC Centre of Excellence for Mathematical and Statistical Frontiers.} \\ School of Mathematical and Physical Sciences\\ University of Technology Sydney\\ NSW 2007, Australia}

\maketitle

\begin{abstract}
Cascades of Poisson processes are probabilistic models for spatio-temporal phenomena in which (i) previous events may trigger subsequent events, and (ii) both the background and triggering processes are conditionally Poisson.
Such phenomena are typically ``data rich but knowledge poor'', in the sense that large datasets are available yet a mechanistic understanding of the background and triggering processes which generate the data are unavailable.
In these settings nonparametric estimation plays a central role.
However existing nonparametric estimators have computational and storage complexity $\mathcal{O}(N^2)$, precluding their application on large datasets.
Here, by assuming the triggering process acts only locally, we derive nonparametric estimators with computational complexity $\mathcal{O}(N\log N)$ and storage complexity $\mathcal{O}(N)$.
Our approach automatically learns the domain of the triggering process from data and is essentially free from hyperparameters.
The methodology is applied to a large seismic dataset where estimation under existing algorithms would be infeasible.
\end{abstract}

\maketitle

\section{Introduction}

Several important real-world processes can be conceptualised as a series of events occurring in time and space, possibly spontaneously, where each event has the capacity to trigger subsequent events.
For example, events in epidemiology correspond to the infection of an individual by a transmissible disease; the infected individual may then go on to cause subsequent infections.
In many settings it is possible to obtain large, detailed data that catalogue the occurrence of events but do not specify individual cause-effect relationships.
For example, in epidemiology it is generally not possible to identify the cause of a specific infection.
Cascades of Poisson processes (COPP) have been successfully applied to facilitate inference and prediction in this setting.
COPP were originally developed by the geological community in the 1980s for statistical modelling of earthquake/aftershock data \citep{Ogata}. 
Recent and varied applications of COPP models have included modelling the dynamics of retaliatory gang violence \citep{Mohler}, terrorist activity \citep{Porter,White}, success of commercial book sales \citep{Deschatres}, military conflicts \citep{Blundell}, disease at the cellular level \citep{Sornette2}, ``retweet cascades'' on the social network Twitter and editing patterns on Wikipedia \citep{Simma}.

In mathematical terms, COPP may be viewed as branching processes with immigration.
Inference in general branching processes is challenging since complex, nonlinear, multidimensional models do not generally admit closed form solutions for estimators \citep{Ogata}.
Moreover COPP models can produce multimodal or very flat log-likelihood functions, precluding reliable numerical algorithms.
To overcome these difficulties, \cite{Zhuang} viewed inference as an incomplete data problem and adapted the Expectation-Maximisation (EM) algorithm \citep{Dempster} to this setting.
In brief, the information about which events arose from the background, which events were triggered and what those triggers were (collectively the ``branching structure'') is required to specify the likelihood, but is unobservable.
By describing the branching structure probabilistically, the EM algorithm seeks instead to maximise the expected likelihood, which can be achieved for certain parametric COPP models (see \cite{Veen} for examples).

In the absence of a mechanistic understanding of either the background or triggering processes, nonparametric estimators play an important role \citep{Zhuang,Marsan}.
However \cite{Sornette} showed that these estimators can exhibit large bias at small-to-moderate sample sizes ($N<10^4$); it is therefore important to integrate as many samples as possible into nonparametric inference.
Unfortunately existing algorithms scale poorly, with computational and storage complexity of existing approaches being $\mathcal{O}(N^2)$.
For example, on earthquake data \cite{Marsan} reported using a sample size of $N = 6,109$ due to computational limitations, whereas seismological datasets now regularly exceed $N > 10^5$ \citep{Hutton}. 
The ``big data'' now available in many application areas for COPP models delivers a pressing need to develop scalable nonparametric estimation procedures.

In this paper we develop accelerated nonparametrics for COPP.
Our approach is based on truncation of kernel density estimators (KDEs) to include only terms which are temporally and spatially local to an event of interest.
The domains of the KDEs are chosen adaptively based on nearest neighbour distances, rendering the approach essentially free from hyperparameters.
The complete algorithm, described in Section \ref{methods}, enjoys computational complexity of just $\mathcal{O}(N\log N)$ and requires $\mathcal{O}(N)$ storage.
Using seismological data, we demonstrate in Section \ref{results} that our estimators facilitate the analysis of larger datasets than was previously possible and provide an empirical assessment of performance in this setting.
Section \ref{discuss} closes with a discussion of challenges and extensions to our methodology.

\begin{figure*}[t]
\centering
\includegraphics[width = \textwidth]{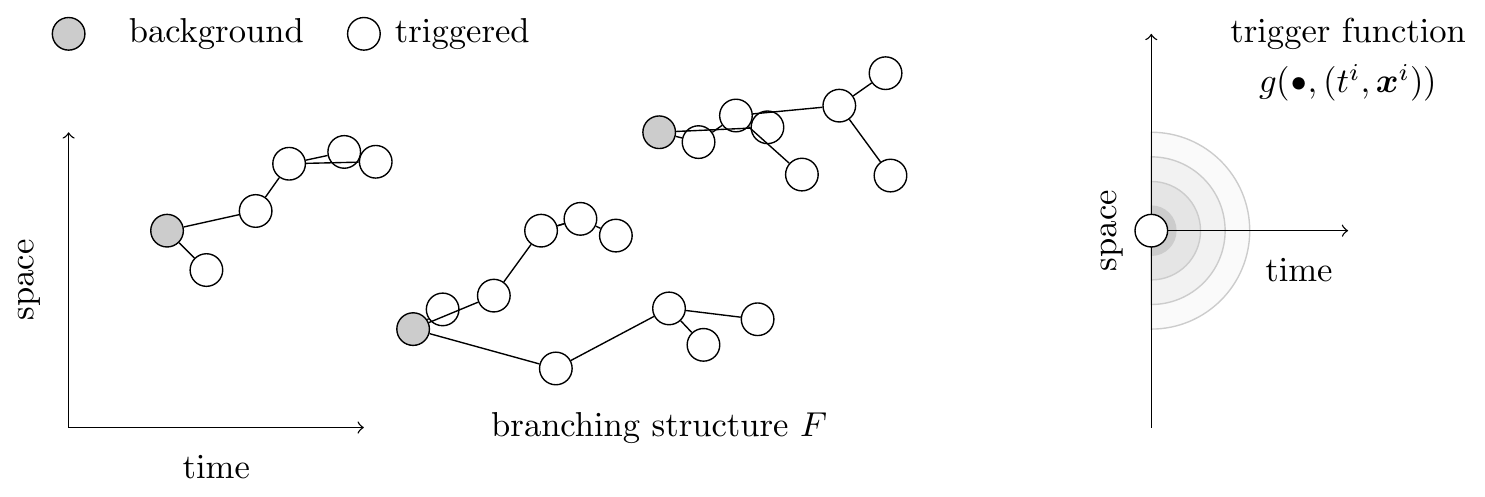}
\caption{An illustration of a cascade of Poisson processes (in $1 \times 1$ dimensions). Left: Background events (grey) arise according to a Poisson process with conditional intensity function $\mu$, whereas triggered events (white) are caused by a previous event $(t^i,\bm{x}^i)$, indicated by an edge in the figure, according to another Poisson process with conditional intensity function $g(\bullet ,(t^i,\bm{x}^i))$. The set $F$ of edges is known as the ``branching structure'' and is statistically equivalent to a random forest. Right: The ``trigger function'' $g$ is strictly zero for $t \leq t_i$, so that events observe a causal order in time. Here contours portray a local, isotropic trigger function, but isotropy is not required for the derivations in this paper.}
\label{illustrate}
\end{figure*}

\section{Methods} \label{methods}

We proceed as follows: Sections \ref{notations} and \ref{defs} set up notation and formally define the COPP model class. Section \ref{algorithms} surveys existing algorithms and nonparametric estimators for COPP. Section \ref{core} contains the core of our methodology for scaling inference to large datasets.

\subsection{Setting and notation} \label{notations}

Consider a metric space $(\mathcal{X},d)$ formed as the finite product $\bigtimes_{i=1}^p (\mathcal{X}_i,d_i)$ where each coordinate $(\mathcal{X}_i,d_i)$ is assumed to be a locally compact metric space.
Here $d$ is a product metric; in this paper all product metrics are induced by the Euclidean norm $\| \cdot \|_2$, so that $d(\bm{x},\bm{y}) = \|(d_1(x_1,y_1),\dots,d_p(x_p,y_p))\|_2$ for $\bm{x},\bm{y} \in \mathcal{X}$.
Consider a simple point process $X$ on $[0,\infty) \times \mathcal{X}$ adapted to a filtration $\mathcal{F}_t$.
For clarity we equip the time domain $[0,\infty)$ with the Euclidean metric, though an arbitrary locally compact metric may be used.
Write $B_{\delta}(\bm{x}) \subset \mathcal{X}$ for the open ball of radius $\delta>0$ centred on $\bm{x} \in \mathcal{X}$.
A measure $\lambda$ on $[0,\infty) \times \mathcal{X}$, known as the conditional intensity of the point process $X$, is given by
\begin{eqnarray}
\lambda(t,\bm{x}) = \lim_{\Delta t, \delta \downarrow 0}  \frac{\mathbb{P} \left\{ \left. \text{exactly one event occurs in } {[t,t+\Delta t) \times B_{\delta}(\bm{x})}  \right| \mathcal{F}_t \right\} }{\Delta t  \times \delta}. \label{limit}
\end{eqnarray}
In this paper $\mathcal{F}_t$ will be the natural filtration; the history up to time $t$, i.e. $\mathcal{F}_t = \{(t^i,\bm{x}^i):t^i<t\}$ where $t^i$ is the time and $\bm{x}^i$ are additional coordinates associated with event $i$ (e.g. spatial coordinates).
It can be shown that the finite-dimensional distributions of $X$ are characterised by the conditional intensity function $\lambda$; in this paper inference and prediction for $X$ is carried out entirely through inference for $\lambda$.
(We assume $X$ is such that the associated measure $\lambda$ is finite on every compact subset of $[0,\infty) \times \mathcal{X}$ and contains no atoms, so that Eqn. \ref{limit} is well-defined.)

\subsection{Cascades of Poisson processes} \label{defs}

This paper concerns the class of point processes $X$ known variously as branching processes with immigration, COPP models \citep{Simma}, ``self-exciting point processes'' \citep{Mohler}, ``epidemic type aftershock sequences'' \citep{Ogata} and ``Hawkes processes'' \citep{Blundell}.
Such models are characterised by a conditional intensity $\lambda$ which is expressed as a superposition of point processes:
\begin{eqnarray}
\lambda(t,\bm{x}) = \mu(t,\bm{x}) + \sum_{i:t^i < t} g((t,\bm{x}),(t^i,\bm{x}^i))
\label{copp eqn}
\end{eqnarray}
Here $\mu$ denotes a (possibly time-varying) background intensity and each $g(\bullet,(t^i,\bm{x}^i))$ represents a ``triggering'' intensity, giving rise to offspring from the trigger event $(t^i,\bm{x}^i)$.
In this formulation, all events $(t^i,\bm{x}^i)$ are able, in principle, to trigger offspring at future times $t > t^i$; the generative model therefore can give rise to ``cascades'' of events, which explains the COPP nomenclature (Fig. \ref{illustrate}).
Estimation for COPP models is equivalent to estimation for the background ($\mu$) and triggering ($g$) intensities.
In an application to earthquake data below, we model a time-independent background intensity $\mu(t,\bm{x}) = \mu(\bm{x})$, but for completeness we present the general case of time-varying background intensities.

\subsection{Stochastic declustering} \label{algorithms}

In applications of COPP models, data $\mathcal{F}_T$ obtained over a time interval $[0,T]$ do not themselves specify which events arose spontaneously, which events were triggered by previous events and what these triggers actually were (collectively the ``branching structure''). Such information can be though of as a forest $F$ on the events in $\mathcal{F}_T$ and is illustrated in Fig. \ref{illustrate}.
Together $\mathcal{D} = \{\mathcal{F}_T,F\}$ can be thought of as complete data; inference for each of $\mu$ and $g$ based on $\mathcal{D}$ is a straight forward density estimation problem.
In applications where the branching structure $F$ is unobservable, inference based only on $\mathcal{F}_T$ can proceed via the EM algorithm \citep{Zhuang}.
In the case where $\mu_{\bm{\theta}}$, $g_{\bm{\theta}}$ are parametrised by $\bm{\theta}$ \citep{Veen}, the EM algorithm proceeds by alternating between taking an expectation over random forests $F$ (``E-step'') and maximising the expected log-likelihood $\mathbb{E}_{F} \log(p(\mathcal{D}|\bm{\theta}))$ over parameters $\bm{\theta}$ (``M-step'').
This procedure, which is sometimes referred to as ``stochastic declustering'', also exists in nonparametric flavours \citep{Zhuang,Marsan}.

Whilst theoretically elegant, stochastic declustering is highly computational; for each event $i$ it is required to exhaustively enumerate all possible triggers of $i$, along with their associated probabilities of being the actual trigger.
Consequently the M-step must be performed with $\mathcal{O}(N^2)$ weighted samples.
To reduce this computational burden, \cite{Mohler} developed a Monte Carlo alternative which approximates the E-step by sampling a single forest from the conditional distribution $F|\mathcal{F}_T,\mu,g$, then substituting this into the complete data likelihood to facilitate the M-step based on only $\mathcal{O}(N)$ unweighted samples.
As a consequence the algorithm does not converge to a single pair of conditional intensities $\mu$, $g$, but instead samples from a set of plausible intensities.
This algorithm has demonstrated good performance in practice \citep{Mohler} but does not currently share the same theoretical guarantees as the EM approach.
Note that the E-step of both existing approaches requires $\mathcal{O}(N^2)$ computational and storage complexities.
In this paper we develop a principled methodology, based on the EM algorithm, that requires only $\mathcal{O}(N \log N)$ computational complexity and $\mathcal{O}(N)$ storage complexity.

\subsection{Accelerated nonparametrics} \label{core}

Below we outline the main component of our proposed methodology, that targets the computational and storage complexity of both the E-step and the M-step in stochastic declustering.
Background events that are distant (either spatially or temporally) to $(t,\bm{x})$ are likely to contribute little to the value of any estimate $\hat{\mu}(t,\bm{x})$ of the background intensity at $(t,\bm{x})$.
Our proposal is therefore to estimate intensity functions using \emph{only} local information. 
A naive approach would be to restrict attention to events within a ball $B_R(t,\bm{x})$ of fixed radius $R$ about the point $(t,\bm{x})$.
However, in areas where background events are sparse, this ball might not contain enough points to enable an accurate approximation of the background intensity at $(t,\bm{x})$.
Noting that the all-nearest-neighbours problem can be solved with time complexity $\mathcal{O}(N \log N)$ and storage complexity $\mathcal{O}(N)$ \citep{Vaidya}, it is computationally appealing to choose $R$ adaptively such that the ball includes a constant number $L$ of nearest neighbours.
Here $L$ should be taken sufficiently large that approximation error, at the level of derived quantities of interest, is negligible.
Full details are provided below.

It will be convenient to relabel the time variable $t$ as the first coordinate, so that our data $\mathcal{F}_T$ can be represented by a $N \times p$ matrix $\bm{X}$ whose entries in the first column $x_{i,1}$ represent the time of the $i$th event, and the remaining columns $x_{i,2}, \dots, x_{i,p}$ represent additional covariates associated with event $i$ (typically spatial coordinates).
Write $\alpha(i,j)$ for the row index in $\bm{X}$ of the $j$th closest event to event $i$ as measured by the standardised Euclidean distance
\begin{eqnarray}
D_{\bm{\sigma}}(\bm{x}_{i,\bullet},\bm{x}_{j,\bullet}) = \sqrt{\left(\frac{x_{i,1}-x_{j,1}}{\sigma_1}\right)^2 + \dots + \left(\frac{x_{i,p}-x_{j,p}}{\sigma_p}\right)^2}
\label{LNN}
\end{eqnarray}
where $\bm{x}_{i,\bullet} = (x_{i,1}, \dots , x_{i,p})$ and $\sigma_k$ is a characteristic length scale for the covariates $\bm{x}_{\bullet , k} = (x_{1,k}, \dots , x_{N,k})^T$ that must be specified.
We adopt the convention that $\alpha(i,1) = i$ and consider $1 \leq j \leq L$.
Our approach is nonparametric and will be presented here using Gaussian kernels.
We will write $\phi(\bm{x}|\bm{m},\bm{\Sigma})$ for the density of a multivariate Gaussian with mean $\bm{m}$ and covariance $\bm{\Sigma}$, evaluated at the point $\bm{x}$.
Then a simple KDE, based on data points $\bm{x}_{i,\bullet} \in \mathbb{R}^p$, is given by
\begin{eqnarray}
\lambda(\bm{x}) \approx \sum_{i=1}^N \phi(\bm{x}|\bm{x}_{i,\bullet}, D_{\bm{\sigma}}(\bm{x}_{i,\bullet},\bm{x}_{\alpha(i,K),\bullet}) \text{diag}(\bm{\sigma})). \label{simple kde}
\end{eqnarray}
Here $K$ is a hyperparameter that controls kernel bandwidth; an optimal choice, in terms of minimising mean square error (MSE), is given by $K \sim N^{4/(4+p)}$, with corresponding MSE $\sim N^{-4/(4+p)}$ \citep[see e.g.][]{Li}.

{\it Remark 1}: The use of a vanishing-tailed kernel implies that influences are considered to act locally.
We note that our methodology below also applies to non-Gaussian kernels, provided that they are local in this sense.

{\it Remark 2}: The characteristic length scales $\bm{\sigma}$ that must be specified will not affect inference when $L$ is sufficiently large and are therefore not hyperparameters {\it per se}.
Indeed, when $L$ is increased we allow more distant events to be considered as possible causes, but these will have negligible contribution to any sensible KDE.

{\it Remark 3}: The simple, illustrative estimate in Eqn. \ref{simple kde} does not respect the COPP model structure in Eqn. \ref{copp eqn}.
Below we accelerate computation in stochastic declustering, directly exploiting the COPP model structure.

\subsubsection{Accelerated E-step}

For accelerated computation the statistical is encoded by a $N \times L$ matrix $\bm{P}$ with entries
\begin{eqnarray*}
P_{i,1} & = & \mathbb{P}[\text{event }i \text{ arose from the background}] \\
P_{i,l} & = & \mathbb{P}[\text{event }i\text{ was triggered by event }\alpha(i,l)].
\end{eqnarray*}
In particular we assume that $\bm{P}$ is a stochastic matrix (unit row sums), i.e. the cause of an event always belongs to its $L$ nearest neighbours. 
In practice it will be necessary to choose $L$ sufficiently large that results are approximately invariant to further increase in $L$.
The matrix $\bm{P}$ characterises the distribution $F| \mathcal{F}_T,\mu,g$ over forests $F$ given data $\mathcal{F}_T$ and intensity functions $\mu,g$.
Indeed, we have that
\begin{eqnarray}
P_{i,1} = \frac{\mu(\bm{x}_{i,\bullet})}{\lambda(\bm{x}_{i,\bullet})}, \; \; \; P_{i,l} = \frac{g(\bm{x}_{i,\bullet},\bm{x}_{\alpha(i,l),\bullet})}{\lambda(\bm{x}_{i,\bullet})} \mathbb{I}\{x_{i,1} > x_{\alpha(i,l),1}\}
\label{PQ}
\end{eqnarray}
where
\begin{eqnarray*}
\lambda(\bm{x}_{i,\bullet}) = \mu(\bm{x}_{i,\bullet}) + \sum_{l = 1}^L g(\bm{x}_{i,\bullet},\bm{x}_{\alpha(i,l),\bullet}) \mathbb{I}\{x_{i,1} > x_{\alpha(i,l),1}\}.
\end{eqnarray*}
By restricting attention to the $L$ nearest neighbours, this accelerated E-step requires $\mathcal{O}(N)$ complexity and storage, compared to the $\mathcal{O}(N^2)$ of existing approaches.

\subsubsection{Accelerated M-step}

Conditional on $\bm{P}$, we define estimates
\begin{eqnarray}
m_k^{\bm{X}} = \frac{\sum_{i=1}^N P_{i,1} x_{i,k}}{\sum_{i=1}^N P_{i,1}}, \; \; \; (\sigma_k^{\bm{X}})^2 =  \frac{\sum_{i=1}^N P_{i,1} (x_{i,k} - m_k^{\bm{X}})^2}{\sum_{i=1}^N P_{i,1}}
\label{eq mx}
\end{eqnarray}
for the mean and standard deviation of the $k$th covariate corresponding to background events.
Fix an integer $K_1$ such that $2 \leq K_1 \leq L$ and write $\beta(i)$ for the index in $\bm{X}$ of the $K_1$th closest background event to event $i$.
$K_1$ will be used to adaptively select an appropriate kernel bandwidth and can be elicited following the optimal $K_1 \sim (\sum_{j=1}^N P_{j,1})^{4/(4+p)}$ rule.
Now $\beta(i)$ is an unknown quantity with uncertainty encoded by the entries in $\bm{P}$; bearing this in mind we can define
\begin{eqnarray}
d_{i}^{\bm{X}} = \mathbb{E}_F[D_{\bm{\sigma}^{\bm{X}}}(\bm{x}_{i,\bullet},\bm{x}_{\beta(i),\bullet})]
\label{eq dX}
\end{eqnarray}
where the expectation is taken over all possible assignments of background and trigger events, weighted according to $\bm{P}$.
Then $d_i^{\bm{X}}$ is an estimate for the (standardised) distance from event $i$ to the $j$th nearest background event.
These values may be obtained exactly \citep{Zhuang} or numerically using Monte Carlo estimation \citep{Mohler}, in the latter case sampling a branching structure $F$ from $F|\mathcal{F}_T,\mu,g$ and then computing $D_{\bm{\sigma}^{\bm{X}}}(\bm{x}_{i,\bullet},\bm{x}_{\beta(i),\bullet})$ directly based on $F$.
In experiments below we took the latter approach.

We estimate the background intensity at event $\bm{x}_{i,\bullet}$ as
\begin{eqnarray}
\mu(\bm{x}_{i,\bullet}) \approx \sum_{j=1}^N P_{j,1} \phi(\bm{x}_{i,\bullet}|\bm{x}_{j,\bullet}, d_j^{\bm{X}}\text{diag}(\bm{\sigma}^{\bm{X}}))  \approx  \sum_{l=1}^L P_{\alpha(i,l),1} \phi(\bm{x}_{i,\bullet}| \bm{x}_{\alpha(i,l),\bullet},d_{\alpha(i,l)}^{\bm{X}}\text{diag}(\bm{\sigma}^{\bm{X}})). \label{mu en long}
\end{eqnarray}
When $L = N$, the second approximation becomes exact, but for $L < N$, the latter expression provides a relaxation of the former, with favourable computational and storage complexity.

The same principle of adaptive truncation based on nearest neighbour distances is used in the differential domain to approximate the triggering function $g$.
We construct a $NL \times p$ matrix $\bm{Y}$ whose entry in the $L(i-1)+l$th row and $k$th column is given by $x_{i,k} - x_{\alpha(i,l),k}$, i.e. the $p$-dimensional vector that joins event $\alpha(i,l)$ to event $i$. 
We refer to these vectors as ``$\Delta$-events''.
Let $\bm{Q}$ be a $NL \times 1$ vector with $L(i-1)+j$th entry $P_{i,j}$ for $j \neq 1$ and $0$ for $j=1$.
Conditional on $\bm{Q}$, we define
\begin{eqnarray}
m_k^{\bm{Y}} = \frac{\sum_{i=1}^{NL} Q_i y_{i,k}}{\sum_{i=1}^{NL} Q_i}, \; \; \; (\sigma_k^{\bm{Y}})^2 =  \frac{\sum_{i=1}^{NL} Q_i (y_{i,k} - m_k^{\bm{Y}})^2}{\sum_{i=1}^{NL} Q_i}
\label{eq my}
\end{eqnarray}
to be the sample mean and standard deviation of $\Delta$-events.

Write $\gamma(i,l)$ for the index in $\bm{Y}$ of the $l$th closest $\Delta$-event to $\Delta$-event $i$, as measured by $D_{\bm{\sigma}}(\bm{y}_{i,\bullet},\bm{y}_{j,\bullet})$ where $\sigma_k$ is the standard deviation of $\bm{y}_{\bullet,k} = (y_{1,k}, \dots , y_{p,k})^T$, with the convention that $\gamma(i,1) = i$.
Also write $\delta(i)$ for the index in $\bm{Y}$ of the $K_2$th closest cause-effect event to the $\Delta$-event $i$.
As before, the optimal bandwidth can be elicited following the $K_2 \sim (\sum_{i=1}^{NL} Q_i)^{4/(4+p)}$ rule.
Now $\delta(i)$ is an unknown quantity with uncertainty encoded by the entries in $\bm{P}$; bearing this in mind we can define
\begin{eqnarray}
d_{i}^{\bm{Y}} = \mathbb{E}_F[D_{\bm{\sigma}^{\bm{Y}}}(\bm{y}_{i,\bullet},\bm{y}_{\delta(i),\bullet})]
\label{eq dY}
\end{eqnarray}
where the expectation is taken over all possible assignments of background and trigger events, weighted according to $\bm{P}$.
These values may again be obtained exactly or numerically using Monte Carlo estimation, and we did the latter.

We estimate the trigger intensity at $\bm{x}_{i,\bullet}$, as contributed by event $\alpha(i,j)$, as
\begin{eqnarray}
g(\bm{x}_{i,\bullet},\bm{x}_{\alpha(i,j),\bullet}) \approx \sum_{l=1}^L Q_{\gamma(L(i-1)+j,l)} \phi(\bm{y}_{L(i-1)+j,\bullet} | \bm{y}_{\gamma(L(i-1)+j,l),\bullet},d_{\gamma(L(i-1)+j,l)}^{\bm{Y}}\text{diag}(\bm{\sigma}^{\bm{Y}})).
\label{eq g}
\end{eqnarray}
Again, when $L=N$, Eqn. \ref{eq g} is exactly the standard KDE based on all data points $\bm{y}_{i,\bullet}$, but for $L<N$, Eqn. \ref{eq g} provides a relaxation of this estimator with favourable computational and storage complexity.

This accelerated M-step has computational complexity $\mathcal{O}(N \log N)$, compared to the $\mathcal{O}(N^2)$ complexity of existing approaches.
Moreover since the $L$-nearest neighbour problem can be solved using $\mathcal{O}(N)$ storage, we also achieve $\mathcal{O}(N)$ storage requirements, compared to the $\mathcal{O}(N^2)$ of existing approaches.

\alglanguage{pseudocode}
\begin{algorithm}[t!]
\caption{Accelerated nonparametrics for cascades of Poisson processes.}
\label{alg}
\begin{multicols}{2}
\begin{algorithmic}[1]
\item[{\it Initialise:}]
\State Compute and cache $\bm{Y}$, $\alpha(\bullet,\bullet)$, $\gamma(\bullet,\bullet)$
\State Initialise $\bm{P}^{(0)}$ and $\bm{Q}^{(0)}$ 
\State{$n \gets 1$, $\epsilon \gets \infty$}
\item[{\it Stochastic declustering:}]
\While{$\epsilon > 10^{-2}$}
\item[{\it M-step:}]
\State Compute $\bm{m}^{\bm{X}}$, $\bm{\sigma}^{\bm{X}}$ using Eqn. \ref{eq mx}
\State Compute $\bm{d}^{\bm{X}}$ using Eqn. \ref{eq dX}
\State Estimate $\mu(\bm{x}_{i,\bullet})$ using Eqn. \ref{mu en long}
\State Compute $\bm{m}^{\bm{Y}}$, $\bm{\sigma}^{\bm{Y}}$ using Eqn. \ref{eq my}
\State Compute  $\bm{d}^{\bm{Y}}$ using Eqn. \ref{eq dY}
\State Estimate $g(\bm{x}_{i,\bullet},\bm{x}_{j,\bullet})$ using Eqn. \ref{eq g}
\item[{\it E-step:}]
\State Re-estimate $\bm{P}^{(n)}$ and $\bm{Q}^{(n)}$ using Eqn. \ref{PQ}
\State $n \gets n+1$, $\epsilon \gets \|\bm{P}^{(n)} - \bm{P}^{(n-1)}\|_1 / N$
\EndWhile
\end{algorithmic}
\end{multicols}
\vspace{-10pt}
\end{algorithm}

The accelerated algorithm presented above, which is somewhat tricky to derive, is the main contribution of this paper.
Complete pseudocode is provided in Alg. \ref{alg}, which proceeds by initialising the distribution over forests, as encoded by $\bm{P} = \bm{P}^{(0)}$. 
For all experiments in this paper we used a uniform distribution such that $P_{i,0} = 1/2$ and $P_{i,j} = 1/(2L)$ otherwise.
Then entries of $\bm{P}$ were set to zero according to whether $x_{i,1} > x_{j,1}$ and $\bm{P}$ was row-normalised to ensure each row defines a probability distribution.
The algorithm is terminated when consecutive iterations change the distribution over branching structures $F$ by less than $10^{-2}$ in total variation distance.

{\it Remark 4}: As is common for EM-type algorithms, formal convergence analysis is mathematically intractable; the estimator need not converge to a global optimum and estimator performance must be assessed empirically.

\section{Results} \label{results}

We proceed as follows: Section \ref{eqdata} describes a typical application of COPP models arising in seismology. Section \ref{AT sec} investigates empirically the computational advantages of the accelerated methodology in this setting. Finally Section \ref{geo sec} compares data-driven nonparametric estimation with model-based estimation via predictive likelihood scores.

\subsection{Earthquake data} \label{eqdata}

We obtained data on $N = 564,750$ earthquakes occurring in a rectangular area around Los Angeles between longitudes 122$^{\circ}$W and 144$^{\circ}$W and latitudes 32$^{\circ}$N and 37$^{\circ}$N (733 km $\times$ 556 km) between January 1st, 1932 and December 31th, 2012  \citep{Hutton}.
Data, which are available for download from http://www.data.scec.org/, include occurrence times and locations based on measurements from $\approx 400$ sensors positioned throughout Southern California.

Decades of geophysical research have led to a deep understanding of the statistical properties of earthquake aftershocks \citep{Ogata}. 
A widely used parametrisation for the trigger function is the ``epidemic-type aftershock model'' 
\begin{eqnarray}
g_{\bm{\theta}}(\Delta t, \Delta \bm{x},m_i) = \frac{K_0 e^{a(m_i-M_0)}}{(\Delta t + c)^{1+\omega}( \|\Delta\bm{x}\|_2^2 + d )^{1+\rho}}. \label{geology}
\end{eqnarray}
Here the $\Delta$ prefix denotes coordinates relative to the $i$th event $(t^i,\bm{x}^i)$ and $m_i$ is its associated magnitude.
Based on Eqn. \ref{geology}, \cite{Veen} performed inference for $\bm{\theta} = \{K_0,a,c,\omega,d,\rho\}$ using a subset of $N = 6,796$ events from the Southern California dataset post-1984 (where data are considered complete above magnitude $M_0=3$), based on a piecewise constant partition of $\mathcal{X} = \mathbb{R}^2$ according to 8 regions of geological fault activity.
Given that the approach of \cite{Veen} is heavily constrained by Eqn. \ref{geology}, it is compelling to see whether nonparametric, data-driven models can compete with this parametric benchmark.
Our nonparametric estimators are constructed as mixtures of Gaussians and are able, in principle, to approximate non-radial intensities such as Eqn. \ref{geology} with arbitrary precision.
Previously it had not been possible to perform this comparison, since nonparametric estimators were computationally restricted to $N \approx 10^3$ samples, which \cite{Sornette} argued was insufficient for robust estimation.

In experiments below we follow the seismology literature by assuming a time-independent background intensity $\mu(t,\bm{x}) \equiv \mu(\bm{x})$.
We did not include magnitude as a coordinate of $\mathcal{X}$, since aftershock magnitude need not be similar to mainshock magnitude.
Characteristic length scales, required for defining the $L$ nearest neighbours in Eqn. \ref{LNN} but not hyperparameters {\it per se}, were taken to be 1 day, 0.1$^{\circ}$ latitude and 0.1$^{\circ}$ longitude.
Boundary effects were not modelled.

\subsection{Accelerated nonparametrics} \label{AT sec}

\begin{figure}[t]
\centering
\includegraphics[width = \textwidth]{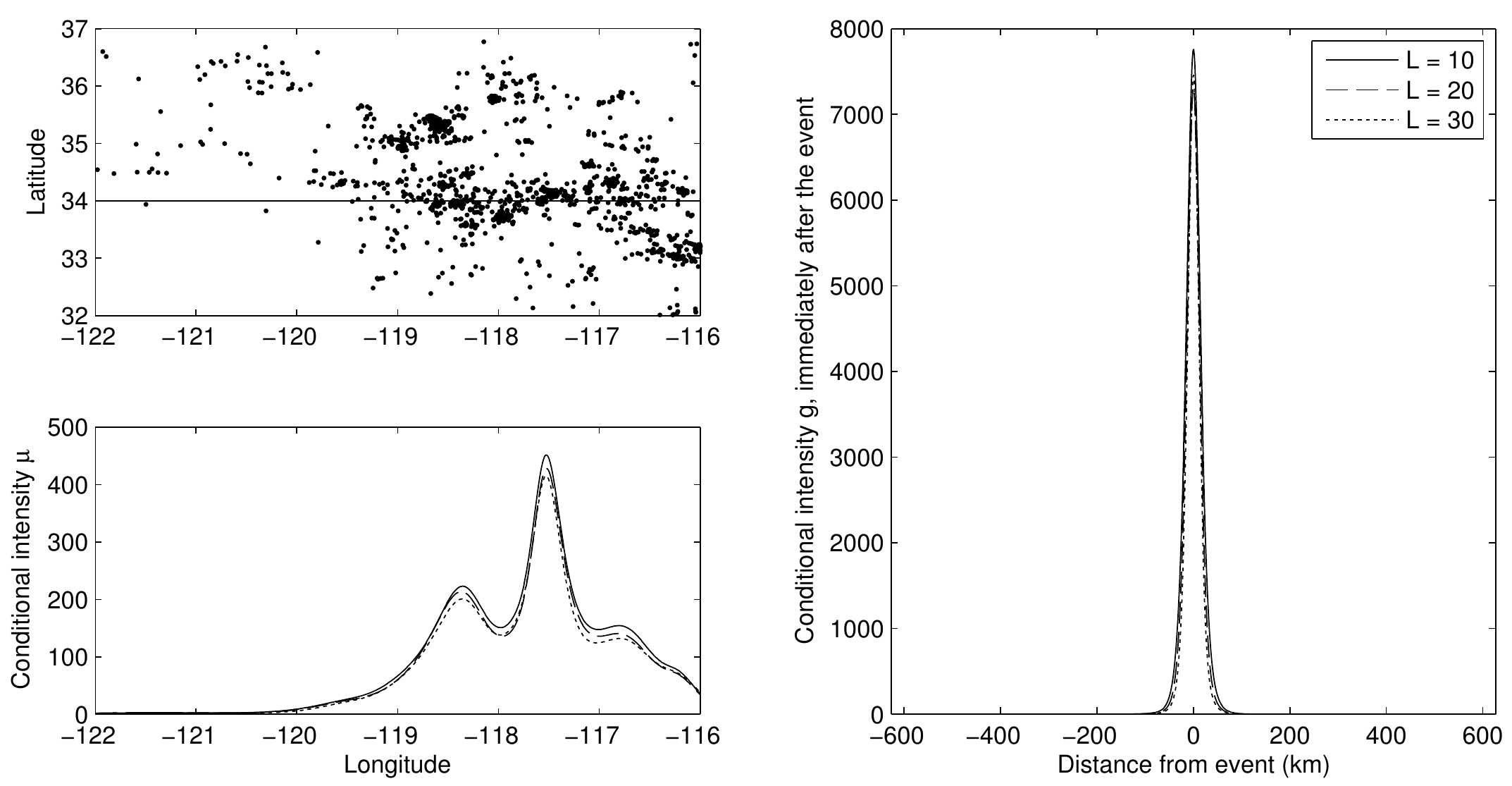}
\caption{Selecting the adaptive truncation parameter $L$. Top: Earthquake data available for the years 1960-1965, with the 34th parallel indicated as a horizontal line. Bottom: Estimated background intensity $\mu$ of earthquakes along the 34th parallel.}
\label{vary L}
\end{figure}

The proposed approach has computational and storage requirements which are linear in the adaptive truncation parameter $L$, with larger $L$ leading to smaller approximation error.
In order to inform our choice of $L$ we plotted the estimated background intensity $\hat{\mu}$ for a hypothetical earthquake along the 34th parallel, varying $L$.
Fig. \ref{vary L} suggests that results, based here on the $N = 1509$ events recorded between 1960 and 1965, are approximately independent of $L$ when $L \geq 10$; we therefore took $L = 10$ for all subsequent experiments.

Our methodology aims to relax current limitations on the size and scope of nonparametric analyses; to test this we implemented both the proposed and existing algorithms on the same platform (MATLAB R2015a) and performed calculations using a single 2.53GHz processor and 3GB of RAM.
Specifically, we compared against exact \citep{Zhuang} and approximate \citep{Mohler} stochastic declustering.
To ensure fair comparison, all algorithms were based on the same Gaussian kernel with fixed (non-adaptive) bandwidth parameters $K_1 = K_2 = 10$ and identical total variation stopping rules were used.
The sampling-based algorithm of \cite{Mohler} does not converge to a unique estimate; we therefore terminated this algorithm after 10 iterations in situations where convergence in total variation was not achieved.
The threshold 10 was chosen since our accelerated estimator typically required fewer than 10 iterations to converge in the above sense.

\begin{figure}[t]
\centering
\includegraphics[clip,trim = 0.1cm 0.1cm 0cm 0.1cm,width = 0.55\textwidth]{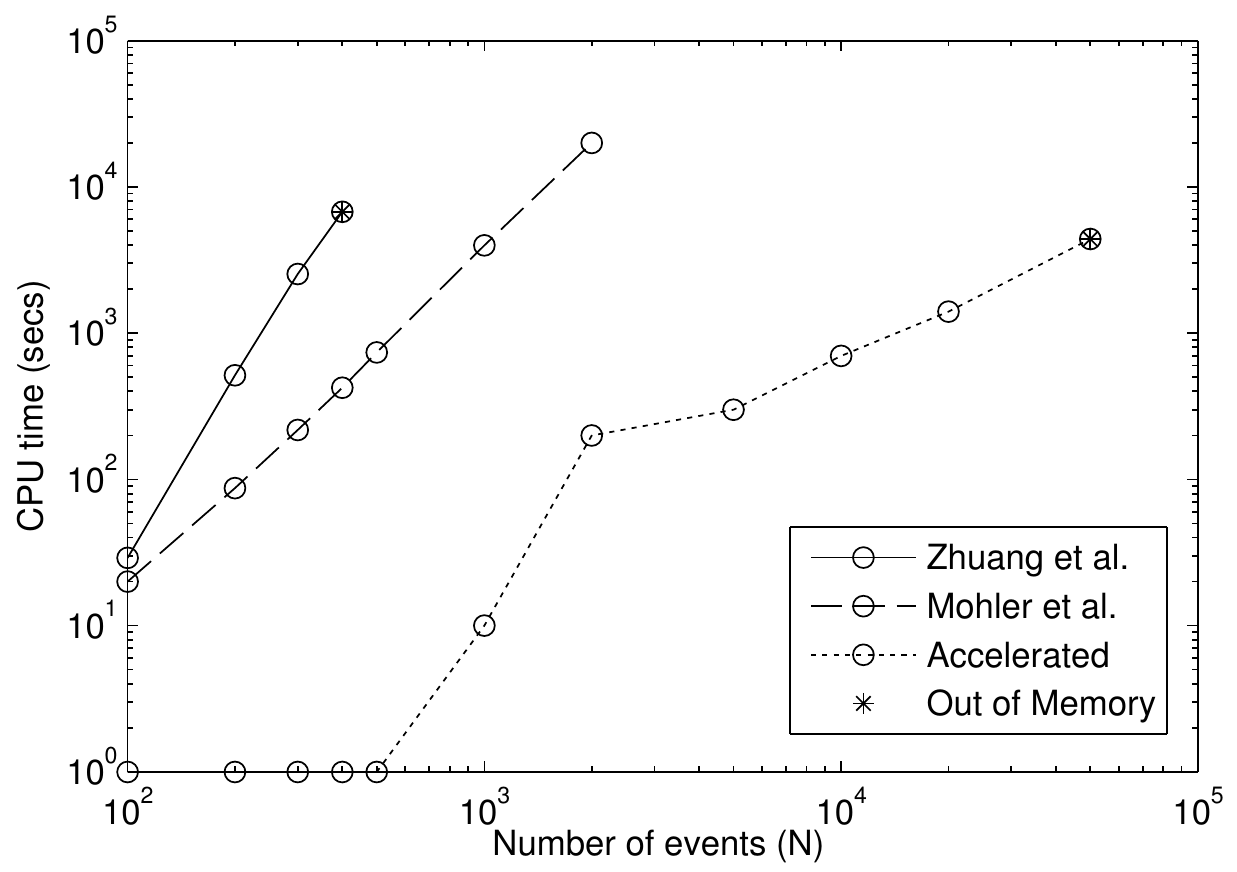}
\caption{Computational times. [Here asterisks are used to denote the final point at which execution of the algorithm was permissible under storage limitations. The proposed ``Accelerated'' approach returned an ``Out of Memory'' error at $N = 10^5$.]}
\label{times}
\end{figure}

A dataset was constructed based on the first $N$ events from the starting point of January 1st, 1932.
We examined the computational time required for termination of each of the three algorithms, while increasing $N$. 
Fig. \ref{times} demonstrates that our accelerated approach is significantly quicker than both exact \citep{Zhuang} and approximate \citep{Mohler} stochastic declustering.
The method of \cite{Zhuang} was heavily constrained by $\mathcal{O}(N^2)$ storage and quickly ran out of memory, whereas \cite{Mohler} was able to go further, with storage $\mathcal{O}(N)$, but was limited by CPU time.
In contrast, the proposed methodology is able to quickly scale to much larger sample sizes.
We emphasise that, whilst it is surely possible to improve each implementation for a given $N$, it remains true that the proposed methodology enjoys favourable computational $\mathcal{O}(N \log N)$ and storage $\mathcal{O}(N)$ complexities, with negligible loss of accuracy compared to the $\mathcal{O}(N^2)$ competing approaches.

\subsection{Parametric versus nonparametric} \label{geo sec}

\begin{figure}[t]
\centering
\includegraphics[width = 0.7\textwidth]{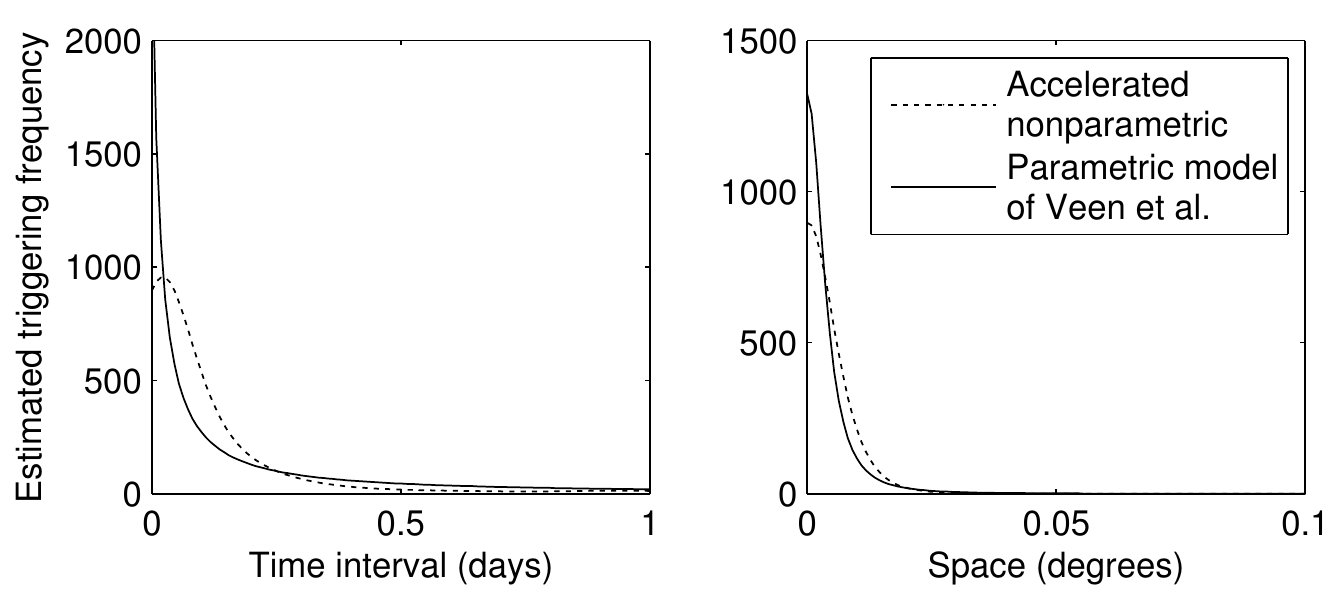}
\caption{Comparison of triggering frequencies, proportional to $\hat{g}$, inferred using the 1984 data and (a) the parametric method of \cite{Veen}, which uses geological knowledge, and (b) the proposed nonparametric methodology. [The parametric estimator depends on the magnitude $m_i$ of the mainshock; for visualisation we chose a magnitude which equates total area under both curves.]}
\label{g}
\end{figure}

Our accelerated methodology allows us to investigate, for the first time, whether nonparametric estimators based on large datasets can be competitive with domain-specific parametric models.
We initially took the parametric model of \cite{Veen}, based on decades of geological research, as a proxy for the true data-generating intensities in the Southern California dataset.
In contrast no geological knowledge entered into the nonparametric estimators.
We then compared our nonparametric estimator $\hat{g}$ against the parametric $g_{\bm{\theta}}$ of Eqn. \ref{geology}, where the former was based on the $N = 17,891$ events occurring in 1984 and the latter based on parameters $\bm{\theta}$ reported in \cite{Veen}.
Fig. \ref{g} shows that $\hat{g}$ approximately recovered the correct support of $g_{\bm{\theta}}$, with the spatial marginal being more accurate than the time marginal.
This rough agreement give confidence that the proposed nonparametric estimators are indeed targeting the correct data-generating intensities. 
In order to probe robustness of these conclusions, we repeated the procedure using data on years 1985-1990; in each case a similar level of approximation was observed between $\hat{g}$ and $g_{\bm{\theta}}$ (see Fig. S1).

Encouraged by accurate recovery of the trigger function, we then assessed the predictive performance of nonparametric methods.
The standard approach to testing earthquake models was established by the working group for the development of Regional Earthquake Likelihood Models (RELM) in 2001 and is reviewed in \cite{Bray}.
In brief, each competing model is required to estimate the number of earthquakes in each of a number of spatio-temporal bins, where the number of events in each bin is assumed to follow a Poisson distribution with intensity parameter equivalent to the forecasted rate.
The simplest performance measure in this setting is known as the L-score, that evaluates the joint probability of held-out data according to the proposed model, computed as a product of independent Poisson probabilities.
Using both our accelerated nonparametric estimator and the parametric model of \cite{Veen}, we attempted to predict earthquakes for each of the 31 days of December in each of 2010, 2011 and 2012, given the previous 7 days' events. 
Our nonparametric approach was based on a large dataset containing the $N = 90,601$ events recorded from 2003 to 2009.
Given estimated intensity functions, the 7 days prior to each day in December were used to construct a predictive intensity over the domain of the held-out data.
By computing the predicted number of events occuring in each of the 30 regions whose boundaries are defined by integer values of latitude and longitude, we are able to quantify predictive performance under the L-score, such that larger values represent better performance.
Results showed that accelerated nonparametric methods were competitive with, but not superior to, model-based prediction ($\log\text{-L} = 3.34$ versus $\log\text{-L} = 3.92$ respectively).
Deconstructing this result, we found that nonparametric methods tended to systematically under-estimate the reproductive ratio $R$ (the expected number of offspring from any given event) relative to the parametric estimator. 
Fig. S2 compares estimates for $R$ based on data from each of the years 1984-1990.
Due to under-estimation of $R$, more events were deemed to be background and were not foreseen, explaining the lower L-scores.

\begin{figure}[t]
\centering
\includegraphics[width = 0.8\textwidth]{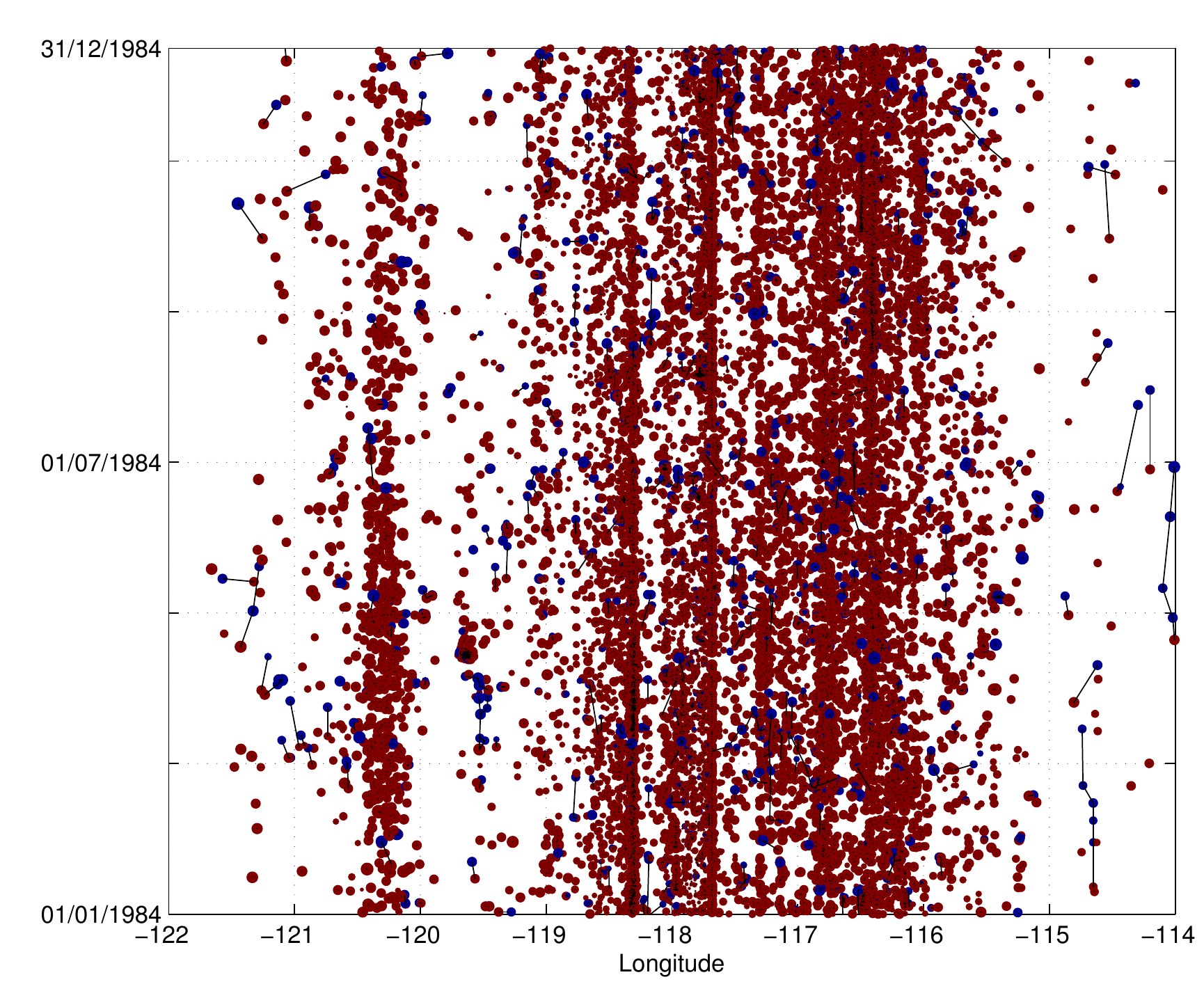}
\caption{A typical sample from an inferred distribution over branching structure, or forest $F$, for earthquake data from 1984 with latitude between 32$^{\circ}$N and 37$^{\circ}$N. [Each point corresponds to one recorded seismic event. Edges are used to join mainshocks (red) to their aftershocks (blue). Size corresponds to earthquake magnitude, which was not used here to estimate the branching structure. To improve visualisation, we do not display latitude information in this figure.]}
\label{forest}
\end{figure}

In addition to estimation of intensity functions, in principle one can also estimate the branching structure $F$.
Fig. \ref{forest} displays a typical point estimate for branching structure.
However, identification of $F$ from occurence data is fundamentally extremely challenging, as pointed out by \cite{Sornette}.
Indeed, we observe in Fig. \ref{forest} that $F$ contains several mainshock-aftershock links that correspond to a time delay of several weeks; this would typically be considered unrealistic on geological grounds and supports the intuition that it is extremely challenging to achieve accurate estimation of branching structure.

\section{Discussion} \label{discuss}

Point processes that admit COPP structure arise in many topical scientific analyses, where typically a precise understanding of the background ($\mu$) or triggering ($g$) processes is currently unavailable.
For example it is unclear how to formulate a parametric model for the triggering process underlying crime waves, or for the spread of infection through a human population.
The nonparametric methods described here have potential to provide new insights in such systems.
In this contribution we accelerated computation for these estimators:
Using adaptive truncation based on nearest neighbour distances, we were able to attain computational complexity $\mathcal{O}(N \log N)$ and storage complexity $\mathcal{O}(N)$, with negligible loss of accuracy compared to existing $\mathcal{O}(N^2)$ procedures.
Using seismology data as a test-bed we demonstrated a practical increase in algorithmic efficiency that allowed the integration of more data for fixed computational cost. 

The efficiency of our approach resulted from adaptive truncation in domains of both the background and trigger intensity functions.
A non-adaptive truncation was previously proposed in \cite{Simma2}.
There an absolute threshold $\tau$ in time was applied, beyond which the triggering function was not evaluated, assumed to be zero. 
In that approach, $\tau$ must be chosen by hand, which could be difficult in settings where little is known about the triggering process.
Our proposal, in comparison, thresholds not only in the time domain but also in the space domain and the domain of the background intensity function.
The resulting computational complexity is $\mathcal{O}(N\log N)$ with constant $C$ proportional to the average number of events occurring in the volume $[t-\tau,t] \times B_{\delta}(\bm{x})$ for thresholds $\tau,\delta$.
Moreover, unlike \cite{Simma2}, our methodology provides a mechanism to select $\tau,\delta$ adaptively, by implicitly solving for $C = L$, thereby mitigating an important practical issue.
The truncation parameter $L$ is not a hyperparameter {\it per se} and should be chosen sufficiently large that any further increase in $L$ leads to negligible variation in the estimated intensity functions, or indeed any derived quantities of interest.

Our preliminary empirical investigation demonstrated inferential and predictive performance that was competitive with parametric estimators, but also revealed systematic downward bias in estimation of reproductive ratios $R$. 
This may be because the introduction of nonparametric uncertainty into the trigger function raises the evidence threshold to conclude that an event was triggered.
Further research will be required to address this methodological issue; indeed, this contribution suggests a number of interesting extensions that are made possible by accelerated computation; (i) reformulating the proposed estimators within the Bayesian framework, allowing for (a) sequential updating of estimators $\hat{\mu}$, $\hat{g}$ as new data arrive, and (b) regularising the reproductive ratio $R$ via a prior distribution $p(R)$, (ii) incorporating observation noise into KDE, (iii) introducing latent variables to account for missing occurrence data, and (iv) migrating nearest neighbour computation to GPUs \citep{Pan}.

Whilst we focused on the popular class of processes with continuous state space $\mathcal{X}$, there exist a number of additional techniques to reduce computational complexity in discrete state spaces (e.g. defined by networks); see \cite{Simma,Zhou} for details.

\vspace{20pt}

\noindent {\bf Supporting Information:} Additional information for this article is available. This includes Appendix S1 (comparison with parametric models) and Appendix S2 (reproductive ratio bias).

\end{document}